\documentclass[a4paper,11pt]{article}
\usepackage{jcappub} 
\arxivnumber{2311.17062} 
\usepackage{amsfonts}
\usepackage{amsmath}
\usepackage{amssymb}
\usepackage{bm}
\usepackage{float}

\newcommand{\hi}{[H {\footnotesize I}]$_{\text{21cm}}$ }
\newcommand{\cii}{[C {\footnotesize II}]$_{158\mu\rm m}$}
\newcommand{\oiii}{[O {\footnotesize III}]$_{88\mu\rm m}$}

\title{Impact of astrophysical scatter on the Epoch of Reionization [H {\huge I}]$_{\text{21cm}}$ bispectrum}

\author[a]{Chandra Shekhar Murmu,}
\author[b]{Kanan K. Datta,}
\author[a,c]{Suman Majumdar,}
\author[d,e,f]{Thomas R. Greve}

\affiliation[a]{Department of Astronomy, Astrophysics and Space Engineering, Indian Institute of Technology, Indore 453552, India}
\affiliation[b]{Relativity \& Cosmology Research Centre, Department of Physics, Jadavpur University, Kolkata 700032, India}
\affiliation[c]{Department of Physics, Blackett Laboratory, Imperial College, London SW7 2AZ, UK}
\affiliation[d]{Cosmic Dawn Center (DAWN), Denmark}
\affiliation[e]{National Space Institute, DTU Space, Technical University of Denmark, Elektrovej 327, DK-2800 Kgs. Lyngby, Denmark}
\affiliation[f]{Department of Physics and Astronomy, University College London, Gower Street, London WC1E 6BT, UK}

\emailAdd{chandra0murmu@gmail.com}

\abstract{It is believed that the first star-forming galaxies are the main drivers of cosmic reionization. It is usually assumed that there is a one-to-one relationship between the star formation rate (SFR) inside a galaxy and the host halo mass in semi-analytical/numerical modeling of large-scale ionization maps during the epoch of reionization. However, more accurate simulations and observations suggest that the SFR and ionizing luminosity in galaxies may vary considerably even if the host halo mass is the same. This astrophysical scatter can introduce an additional non-Gaussianity in the \hi signal, which might not be captured adequately in the power spectrum. In this work, we have studied the impact of the scatter on the \hi bispectrum using semi-numerical simulations. We find that the scatter primarily affects small ionized regions, whereas the large ionized bubbles remain largely unaffected. Although the fractional change in the \hi bispectra due to the scatter is found to be more than a factor of $10$ at large scales ($k_1 \lesssim 1\, {\rm Mpc}^{-1}$) for $z=7.4$, it is found to be statistically insignificant. However, at small scales ($k_1\sim2.55$~Mpc$^{-1}$), we have found the impact due to the scatter to be high in magnitude ($|\langle \Delta B \rangle/B_{\text{no-scatter}}| \sim 1$) and statistically significant ($|\langle\Delta B\rangle/\sigma_{\Delta B}| \gtrsim 5$) at neutral fraction, $\overline{x}_{\rm HI}\sim 0.8$ for $z=7.4$. The impact due to scatter is found to be even more prominent ($|\langle \Delta B \rangle/B_{\text{no-scatter}}| \gtrsim 10$) at small scales for $z=10$ and $\overline{x}_{\rm HI}\sim 0.8$, but with reduced statistical significance to some extent ($|\langle\Delta B\rangle/\sigma_{\Delta B}| \sim 3$), compared to $z=7.4$ at the same neutral fraction. We have also found that in the most optimistic scenario, SKA1-Low might be able to detect these signatures of astrophysical scatter, at $\sim 3\sigma$ and $\sim 5\sigma$ detection significance for $\overline{x}_{\rm HI} \sim$ 0.8 and 0.9 respectively, for the equilateral \hi bispectrum at $z=7.4$.}


\begin{document}
\maketitle
\flushbottom

\section{Introduction}
\label{sec:intro}
Modeling observable summary statistics for line-intensity mapping (LIM) signal is one of the key steps to understand the poorly constrained epoch of reionization (EoR). Line emissions from either the IGM or the galaxies provide us the opportunity to map the Universe and its cosmic evolution.
The LIM signal corresponding to the redshifted \hi emissions from the diffuse IGM~\citep{Wouthuysen_1952, Field_1958, Madau_1997}, have been proposed as an excellent probe to track the epoch of reionization. On the other hand, independent and complementary probes using bright line emissions from the galaxies such as \cii~\citep{Gong_2012, Silva_2015, Padmanabhan_2019, Yue_2019, Murmu_2021, Murmu_2022, Karoumpis_2022}, CO \citep{Lidz_2011, Li_2016, Yang_2022, Dizgah_2024, Roy_2023}, Ly-$\alpha$ ~\citep{Peterson_2012, Silva_2013, Pullen_2014, Mas-Ribas_2020}, \oiii~\citep{Padmanabhan_2023} etc. have also been proposed for studying the EoR. In that process, one can use suitable summary statistics to infer useful astrophysical and cosmological information from these LIM probes. Fourier-based statistics, such as the power spectrum, provide information on signal fluctuations at different length scales. Higher-order statistics, such as the \hi bispectrum, can reveal non-Gaussian features present in the signal at multiple length scales.\par
These statistics are sensitive to various phenomena, such as source properties, star formation rate, line-of-sight effects, astrophysical processes of line emissions, etc. Appropriate inferences of various information require these signal statistics to be modeled accurately. One of the usual approaches in this modeling is to connect the galaxy line emissions to their host halo mass using various scaling relations and predict LIM signals. This approach offers flexibility in simulating LIM signals at large (cosmological) scales relatively quickly, as opposed to more accurate hydro-simulations. Normally, these models assume a one-to-one correspondence between the host halo mass and the line luminosity of interest~\citep{Gong_2011, Lidz_2011, Silva_2013, Breysse_2014, Pullen_2014}. However, in reality, the line luminosity from galaxies can vary due to various astrophysical reasons even if the host halo mass is the same. This affects the summary statistics. The effect of this astrophysical scatter has been explored in the context of galaxy LIM signals, which is shown to enhance the power spectrum at small scales ~\citep{Li_2016, Schaan_2021, Yang_2022}. A more generalized non-uniform line-luminosity scatter affects the large-scale power spectrum as well \citep{Moradinezhad_Dizgah_2022, Murmu_2023}.\par
A similar effect of astrophysical scatter can also be present in the star-formation rates (SFR) of reionizing galaxies. There can be variations in the SFR, even if the host halo mass of a given galaxy is the same. Assuming that the emission rate of ionizing photons from the galaxies is correlated with its SFR, the halo-to-halo scatter in the SFR will affect the number distribution of UV ionizing photons emitted. These photons, in turn, ionize the neutral IGM, and therefore, the scatter would leave imprints on the ionization fluctuations. Consequently, it will affect the cosmological \hi signal emerging from the IGM during the EoR. A study by~\cite{Hassan_2022} explored the role of this astrophysical scatter for the first time in the context of cosmic reionization, using the ionization power spectrum. It was found that the ionization power spectra are mostly unaffected by the presence of scatter. However, the role of this scatter is not well investigated in the context of the observable \hi signal from the EoR. This signal is known to have non-Gaussian features, and the astrophysical scatter might introduce additional non-Gaussianity, which the power spectrum might not capture adequately. On the other hand, one-point statistics such as skewness and kurtosis can capture non-Gaussian signatures in the \hi signal at a particular length scale~\citep{Harker_2009, Shimabukuro_2015, Watkinson_2014, Watkinson_2015, Kubota_2016, Ross_2021}. However, these one-point statistics cannot capture the correlation between multiple-length scales. To characterize the presence of non-Gaussianity in the correlation between different length scales, one needs to resort to higher-order Fourier statistics, such as the bispectrum.\par
The bispectrum is the 3-pt correlation function in Fourier space and correlates three different $\bm{k}$ vectors, which form a closed loop in the Fourier domain. Therefore, different triangle configurations of the $\bm{k}$ vectors can capture the correlation between different length scales, which the power spectrum fails to do. The \hi signal bispectrum is sensitive to non-Gaussian features in the signal and, suitable for analyzing features. Studies by \cite{Majumdar_2018, Majumdar_2020, Hutter_2020, Gill_2024, Raste_2024} have demonstrated that the \hi signal bispectrum can characterize the features of non-Guassianity and topology, which otherwise would be difficult to do with the power spectrum. We describe this in more detail in subsection~\ref{sec:auto-bispectrum}.\par
In this article, we investigate the impact of the astrophysical scatter on the \hi bispectrum. Since the variation in the SFR for a given host halo mass is stochastic, we simulate multiple independent realizations of the \hi maps, at a fixed neutral fraction, by varying the seed of the randomness for the scatter. This is useful in quantifying the statistical significance of the changes in the bispectrum when we take astrophysical scatter into account. We estimate the bispectrum for all unique triangle configurations and for an extensive range of length scales ($k_1$ values). We investigate the length scales and regions of the \hi bispectrum configuration space that are sensitive to the changes introduced by the astrophysical scatter. These changes are also compared with the changes in the power spectrum, which is also computed for the \hi maps with scatter. Finally, we explore the possibility of detecting the bispectrum, at length scales where the changes due to astrophysical scatter are found to be significant, considering the planned SKA1-Low baseline configurations and for various observational scenarios.\par
This paper is organized as follows:
In Section~\ref{sec:scatter}, we discuss the astrophysical scatter. The following Section (Section~\ref{sec:bispec}) summarizes the \hi signal and the bispectrum. The simulation of the \hi maps with astrophysical scatter is described in Section~\ref{sec:sims}. In Section~\ref{sec:res}, we discuss the impact of astrophysical scatter on the bispectrum and the statistical significance of the impact. We also compare this with the impact seen on the power spectrum. In this section, we explore the prospects of detecting the equilateral \hi bispectrum as well, for various observational scenarios with the planned SKA1-Low. Finally, we summarize this work in Section~\ref{sec:summary}. Throughout this work, we have adopted cosmological parameters $\Omega_{\text{m}}=0.3183\,,\Omega_\Lambda=0.6817\,,h=0.6704\,,\Omega_{\text{b}}h^2=0.022032\,,\sigma_8=0.8347\,,n_{\text{s}}=0.9619$, consistent with Planck+WP best-fit values \citep{Planck_XVI}.

\section{Astrophysical scatter}
\label{sec:scatter}
The typical approach for modeling ionizing photon emission from the galaxies is to model their star-formation rates. It is a reasonable approach because star-forming galaxies mainly drive the reionization process. A simple way to model the galaxy SFR is to relate it with the host halo mass. It has been primarily used in the \hi literature to model the reionization process. If we assume $N_{\gamma}$ to be the total number of ionizing photons deposited in the IGM from the instantaneous star formation in a galaxy, then a common model is to assume  $N_{\gamma} \propto M^\alpha_{\text{h}}$. Here, the star formation rate is modeled as a power law of the halo mass, with $\alpha$ being the power law index. Other variants of this model can be a complicated function of the halo mass, usually different power laws at different mass ranges with multiple parameters. We have used one such model in this work, given in Equation \ref{eq:sfr}.\par
However, the stochasticity in these SFR models is usually left out while modeling the cosmic reionization of the IGM and the emanating \hi signal. Observationally, it is found that the SFR of galaxies obeys a tight correlation with the stellar mass. \cite{Speagle_2014} compile data from earlier works \citep{Daddi_2007, Noeske_2007, Magdis_2010, Whitaker_2012}, which used different methods to determine the main-sequence (MS) relation and its dispersion. These were brought to a standard calibration by \cite{Speagle_2014} to obtain the relation between $\log \psi$ -- $\log M_\star$ and the dispersion around it, with $\psi$ being the SFR and $M_\star$ being the stellar mass of the galaxy. For details, interested readers are referred to \cite{Speagle_2014}. They find the true intrinsic scatter is around $0.2$ dexes, with $0.3$ dexes being the upper limit considering observational uncertainties.\par
Various analytical modeling and numerical simulations have studied the origin of this dispersion in the relation. It is understood that the primary driver for this scatter is the varying mass accretion history over time \citep{Dutton_2010, Peng_2014, Matthee_2019, Blank_2021}, which happens over longer time scales. Shorter time scale variabilities, such as short time-scale variabilities in the gas accretion rate, and different feedback mechanisms also contribute and are essential for lower mass galaxies. Various numerical simulations have reproduced the dispersion in the main sequence within a similar range of 0.2--0.3 dexes \citep{Lagos_2018, Matthee_2019, Blank_2021}. Therefore, we adopt the fiducial value of $\sigma=0.3$ in our study, as described in section \ref{sec:sims}. We assumed that this scatter also applies to the high-redshift Universe and that the stellar mass follows a tight correlation with the underlying dark-matter (DM) halo mass.

\section{The [H {\footnotesize I}] 21cm Bispectrum}
\label{sec:bispec}
\subsection{The [H {\footnotesize I}] 21cm signal from the IGM}
The redshifted \hi signal arises from the neutral hydrogen atoms (H{\footnotesize I}) of the Universe via hyperfine spin-flip transition. It is a promising probe of the IGM and the reionization process as it allows us to track the evolution of the Universe through cosmic time. This signal is observed as a differential brightness temperature ($\delta T_{\text{b}}(\bm{x},z)$) against the cosmic microwave background radiation (CMBR), with $\delta T_{\text{b}}(\bm{x},z) \propto x_{\rm HI}(\bm{x},z)(1+\delta_{\text{H}}(\bm{x},z))$ \citep{Bharadwaj_2005}. Here, $x_{\rm HI}$ is the neutral fraction and $\delta_{\text{H}}$ is the baryon overdensity.\par
One of the primary goals is to measure the power spectrum corresponding to \hi signal fluctuations. However, it fails to capture all the information in the signal, specifically if the fluctuations in the signal are not Gaussian-random. The reionization process is non-linear, and the nature of the corresponding fluctuations it introduces in the surroundings will be non-Gaussian \citep{bharadwaj_2005a, Iliev_2006, Mellema_2006, Mondal_2015}. As demonstrated by \cite{Mellema:2015IS}, if we consider a realistic \hi map derived from a numerical simulation and an artificial \hi map with Gaussian-random fluctuations, the power spectrum might fail to distinguish between these two scenarios.\par
Given a fixed halo mass, there will be a log-normal distribution in the number of ionizing photons contributed from those haloes of identical masses. This additional source of non-Gaussianity in the number distribution of ionizing photons is expected to affect the spatial distribution of $x_{\rm HI}(\bm{x},z)$. Keeping the underlying gas density the same, these induced non-Gaussian fluctuations in $x_{\rm HI}(\bm{x},z)$ will reflect in the $\delta T_{\rm b}(\bm{x},z)$, since $x_{\rm HI}(\bm{x},z) \propto \delta T_{\rm b}(\bm{x},z)$. Therefore, under such a scenario, the power spectrum is not expected to entirely capture the impact of scatter on the \hi signal. It motivates us to investigate it using higher-order summary statistics, such as the bispectrum.

\subsection{The auto-bispectrum}
\label{sec:auto-bispectrum}
The auto-bispectrum from numerical simulation can be estimated as,
\begin{equation}
    \hat{B}_m(\bm{k_1},\bm{k_2},\bm{k_3}) = \frac{1}{N_{\text{tri}}V_{\text{box}}}\sum_{[\bm{k_1}+\bm{k_2}+\bm{k_3}=0]\,\in\, m}\Tilde{\Delta}T_{\text{b}}(\bm{k_1})\Tilde{\Delta}T_{\text{b}}(\bm{k_2})\Tilde{\Delta}T_{\text{b}}(\bm{k_3}),
\end{equation}
where $\Tilde{\Delta}T_{\text{b}}(\bm{k})$\footnote{This convention assumes a definition of Fourier transform as $\delta T_{\text{b}}(\bm{x}, z) = \int \frac{d^3k}{(2\pi)^3}\exp{(i\bm{k.x})}\tilde{\Delta} T_{\rm b}(\bm{k}, z)$} is the signal in Fourier space. The three wave vectors $\bm{k_1}$,$\bm{k_2}$, and $\bm{k_3}$ should form a closed loop (a triangle) for a particular $m$ th triangle configuration. The ensemble average of the product of the Fourier quantities is then normalized with the number of triangle configurations, $N_{\text{tri}}$ for the $m$ th bin and the volume of the box, $V_{\text{box}}$. One needs to identify all the possibilities of the triangle configurations in Fourier space to fully characterize the bispectrum. We can parametrize the bispectrum with $n = k_2/k_1$ and $\cos\theta = -\bm{k_2}\bm{.k_1}/(k_2k_1)$, and we can identify all the unique triangle configurations if we label the arms of the triangles such that $k_1 \geq k_2 \geq k_3$ and the conditions $0.5 \leq n, \cos\theta \leq 1.0$ and $n\cos\theta\geq0.5$ are satisfied. For more detail on this characterization, interested readers can refer to \cite{Bharadwaj_2020}.\par
The bispectrum has been studied in detail to explore various characteristics of the \hi signal. One of the features it tries to extract from the \hi signal is the non-Gaussianity, which is mostly lost in the power spectrum. \cite{Majumdar_2018, Majumdar_2020} has investigated the various components of non-Gaussian signal to the \hi field ($x_{\rm HI}$ and $\delta_{\text{H}}$) and how it evolves as reionization proceeds. Further, the sign of the bispectrum has been shown to disentangle the dominance of the contribution of non-Gaussianity from these sources. \cite{Majumdar_2020} and \cite{Gill_2024} have further quantified the impact of the redshift space distortions on this signal statistic and how the non-Gaussianity depends on the signal topology. The studies by \cite{Hutter_2020} and \cite{Raste_2024} have further independently confirmed the dependency of \hi topology on the sign and amplitude of the signal bispectrum.
The existing literature on \hi bispectrum studies of CD-EoR suggests that one needs to forward model the bispectrum to study various effects that can impact the signal from this era. The redshift space distortions (RSD) are shown to significantly impact the EoR \hi signal both in terms of magnitude and sign  \citep{Majumdar_2020, Gill_2024} and affect most of the triangle configuration space. Similarly, the Cosmic Dawn (CD) \hi signal bispectra is affected significantly by the spin temperature fluctuations and the RSD \citep{Kamran_2021a}. The other source of LoS anisotropy, the light-cone effect \citep{datta_2012, zawada_2014, Murmu_2021}, arising due to the finite light travel time of the signal from its sources to the present-day observer, is found to affect the squeezed limit bispectrum above the cosmic variance level \citep{Mondal_2021}. These effects need to be considered for a proper interpretation of the auto-bispectrum. Here, we focus on the impact of the astrophysical scatter on the non-Gaussianity in the \hi signal, in particular on the bispectrum.

\section{Simulating the [H {\footnotesize I}] 21cm maps}
\label{sec:sims}
We have used a combination of N-body dark-matter-only simulation \citep{Bharadwaj_2004} and a semi-numerical prescription for modeling reionization \citep{Choudhury_2009, Majumdar_2014, Mondal_2017}. The side length of the DM simulation box is $215$ Mpc in size, with a total grid number of $3072^3$ and particle number of $1536^3$. We run a FoF algorithm \citep{Mondal_2015} to identify collapsed haloes from the dark-matter distribution snapshots, which are assumed to be the sources of ionizing photons. The emission of ionizing photons is modeled as being proportional to the SFR for a given halo. Following \cite{Silva_2013}, we model the SFR as a function of the halo mass ($M_{\text{h}}$) as,
\begin{equation}
\label{eq:sfr}
\begin{split}
    \frac{\text{SFR}\,(M_{\text{h}}, z)}{M_{\odot}yr^{-1}} = 2.25\times10^{-26}(1+7.5\times10^{-2}\times (z-7))\\
    \times M_{\text{h}}^a\bigg(1+\frac{M_{\text{h}}}{c_1}\bigg)^b\bigg(1+\frac{M_{\text{h}}}{c_2}\bigg)^d\bigg(1+\frac{M_{\text{h}}}{c_3}\bigg)^e,
\end{split}
\end{equation}
with $a=2.59$, $b=-0.62$, $d=0.4$, $e=-2.25$, $c_1=8\times10^8 M_\odot$, $c_2=7\times10^9M_\odot$, and $c_3=1\times10^{11}M_\odot$. Therefore $N_\gamma \propto \overline{\text{SFR}} (M_{\text{h}},z)$, where we assume the SFR to be correlated to the host halo mass through this mean relationship. To introduce scatter, we assume a log-normal distribution for the SFR, implemented as follows:
\begin{equation}
\label{eq:scatter}
    \log \text{SFR}^{\text{scatter}} = \bigg(\log \overline{\text{SFR}} - \frac{1}{2}\sigma^2\ln (10)\bigg) + \mathcal{N} (0, \sigma^2),
\end{equation}
with $\mathcal{N}(0,\sigma^2)$ being the normal distribution with zero mean and standard deviation $\sigma$. Therefore SFR$_\text{scatter}$ has a spread of $\sigma$ dex across the halo mass range. We fix the value of $\sigma$ to 0.3 dex for our entire exercise. The number of ionizing photons emitted in the scatter scenario from a given halo is then $N^{\text{scatter}}_\gamma \propto \text{SFR}^{\text{scatter}}$.\par
The reionization process is simulated over a coarse-gridded box of $384^3$ grids, resulting in a grid resolution of 0.56 Mpc. \hi maps are generated using excursion set formalism \citep{Furlanetto_2004}, where the ionization condition is met when $\langle n_\gamma \rangle_R \geq \langle n_{\rm H} \rangle_R$ for a region smoothed over radius $R$, where $\langle n_\gamma \rangle_R$ and $\langle n_{\rm H} \rangle_R$ are the average number of ionizing photons and hydrogen atoms within a spherical region of radius $R$. Otherwise, an ionized fraction value $x_{\rm HI} =  \langle n_\gamma \rangle_{\text{grid}}/\langle n_{\rm H} \rangle_{\text{grid}}$ is assigned to the grid. This semi-numerical approach of simulating the reionization does not consider density-dependent recombination and we also note that this model is not photon conserving \citep{choudhury_2018}. The impact of photon conservation on the \hi bispectrum has not been studied. However, it has been shown that photon conservation boosts the \hi power spectrum in all scales (without significant change in its shape) and results in a comparatively rapid reionization \citep{choudhury_2018}. The density-dependent recombination is expected to introduce an additional scatter in the \hi signal topology. Both of these may have a significant impact on the \hi bispectrum but their study is beyond the scope of this article. We plan to take up that investigation in a future follow-up work.\par

We generate the \hi maps across an extensive neutral fraction range of $\overline{x}_{\rm HI}$ = [0.53, 0.62, 0.72, 0.81, 0.9, 0.95] at $z=7.4$. As argued by \cite{Hassan_2022}, the presence of ionized bubbles of sufficient size and a significant number will wash away any signatures of scatter. Therefore at high neutral fractions, we investigate whether any signatures of the scatter can be captured using the bispectrum. Since scatter will impact only the $x_{\rm HI}(\bm{x},z)$ field and non-Gaussianity contributions come from $\delta_{\text{H}}(\bm{x},z)$ as well, we fix the redshift for all the neutral fractions to exclusively study the impact of scatter on the bispectrum by removing the contribution of $\delta_{\text{H}}(\bm{x},z)$ to non-Gaussian features. We have simulated $\sim 50$ statistically independent realizations of \hi maps by using $\sim 50$ different seeds of randomness for the scatter at $z=7.4$ for each of the six neutral fractions to estimate the bispectrum. We note here that according to the latest results of Planck 2018~\cite{Planck_XVIII}, the constraint on the mid-point of reionization is $z_{\rm reion} = 7.68 \pm 0.79$ (Eq. 18 of ~\cite{Planck_XVIII}) under the assumption of a reionization history that follows a tanh model. Given the uncertainty on the value $z_{\rm reion}$, it may be possible to have large neutral fraction values at $z=7.4$, as explored in this study. Nevertheless, we have also considered a scenario where the neutral fraction is $\overline{x}_{\rm HI} \approx 0.8$ at a higher redshift of $z=10$, which can be safely regarded as consistent with the Planck 2018 constraints. In this case, we also use the same set of ~50 random seeds to generate multiple realizations of the scatter scenario and estimate the impact of the astrophysical scatter and its corresponding statistical significance. Therefore a total of $\sim 350$ simulations of the \hi maps combining all scenarios were done, which consumed a significant amount of computational time. It will affect how much ionizing flux a given halo will generate under this stochastic model from realization to realization. It helps us to asses any given statistic's variance under the scatter model across the realizations.

\section{Results}
\label{sec:res}
\begin{figure*}
    \centering
    \includegraphics[width=\textwidth]{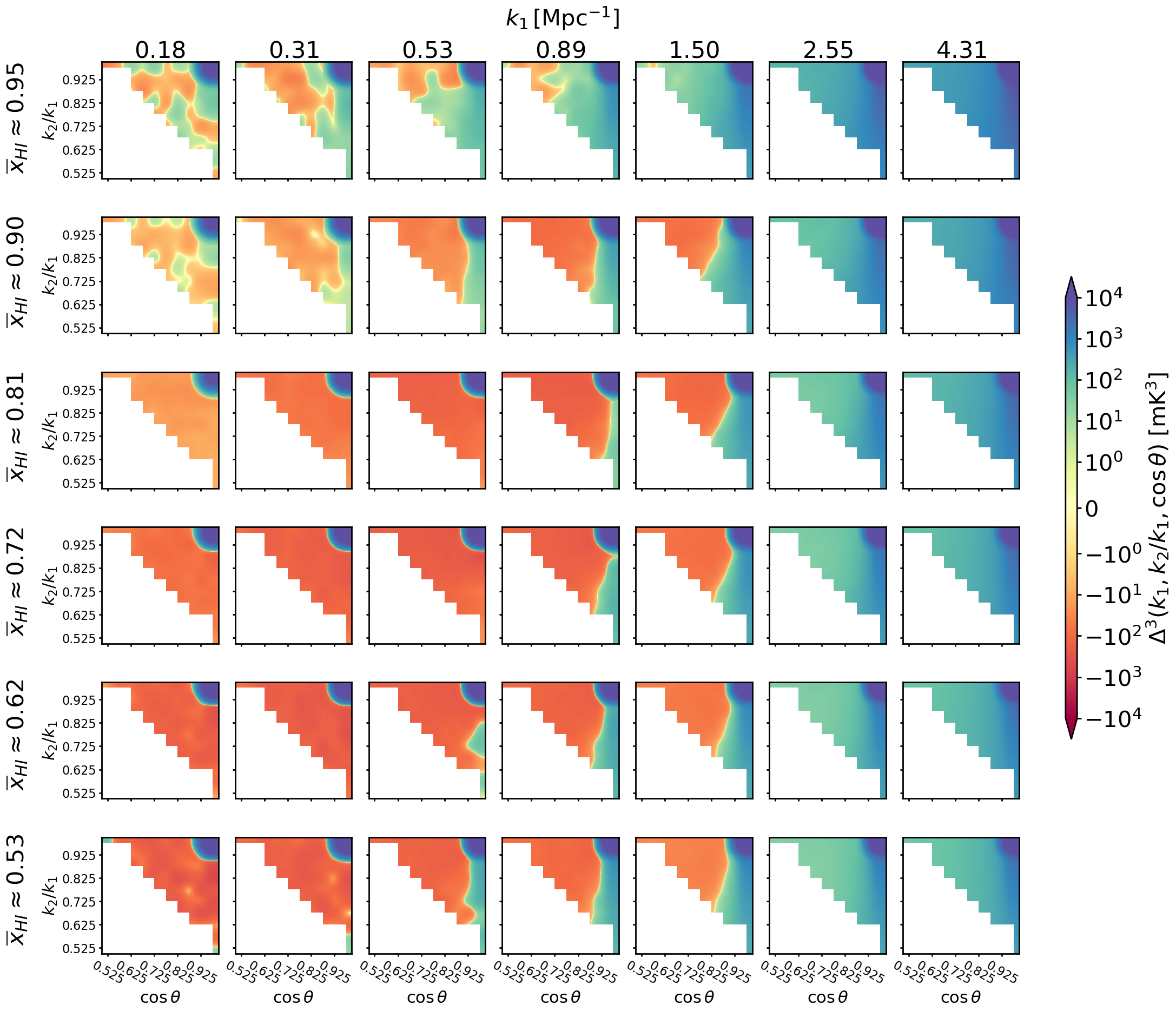}
    \caption{This figure shows the \hi auto-bispectrum for all unique triangle configurations and averaged over all realizations of astrophysical scatter at $z=7.4$. This is shown for the full set of neutral fraction range that has been considered here.}
    \label{fig:bispec_scatter}
\end{figure*}

We used a direct estimator of the bispectrum, as described in \cite{Mondal_2021}. The bispectrum is estimated at a fixed neutral fraction for each realization and then averaged over all realizations. This same exercise is repeated for all the neutral fractions, considered in this study. $\langle B_{\text{scatter}} \rangle$ represents the mean bispectrum estimated from all realizations of \hi maps with astrophysical scatter, for a given neutral fraction. Impact of the astrophysical scatter is quantified by calculating $\langle \Delta B \rangle = \langle B_{\text{scatter}} \rangle - B_{\text{no-scatter}}$ and the modulus of the ratio between $\langle\Delta B\rangle$ and $B_{\text{no-scatter}}$, i.e. $|\langle \Delta B \rangle/B_{\text{no-scatter}}|$, where $B_{\text{no-scatter}}$ is the bispectrum of the \hi map without any scatter. This tells how much the bispectrum, averaged over multiple realizations of astrophysical scatter deviates from the original bispectrum without the impact of scatter. In Figure~\ref{fig:bispec_scatter}, we show the averaged bispectrum (averaged over $50$ statistically independent realizations of the scatter at each neutral fraction) for the \hi maps with scatter. The mean neutral fraction is fixed along the row (labeled on the left), and the $k_1$ value (mentioned on the top) is fixed along the column. Since the $k_1$ value changes along the row,  plots along a row reflect the effect due to the changing size of the bispectrum triangle configuration. The shape of the configuration is parametrized by $\cos\theta$ and  $k_2/k_1$ ratio for a given $k_1$, as mentioned in subsection~\ref{sec:auto-bispectrum}. Therefore, $k_1$ represents the overall length scales over which the correlation occurs between the different $\bm{k}$ vectors.
The results on the impact of scatter on the bispectrum are presented in subsection \ref{impact_on_unique_tri}. In subsection~\ref{stat_sig}, we discuss at what scales the impact on the \hi bispectrum is sufficiently high in magnitude and statistically significant. In subsection~\ref{pow_spec}, we try to understand how the impact of scatter on the power spectrum, which is a more common statistic, compares to that on the \hi bispectrum. Finally, we explore various scenarios for detecting the equilateral \hi bispectrum, which is one of the triangle configurations expected to capture the signatures of scatter, with the planned SKA1-Low, in subsection~\ref{bispec_detect}.

\subsection{Impact of scatter on the bispectrum for all unique triangles}
\label{change_scatter}
\begin{figure*}
    \centering
    \includegraphics[width=\textwidth]{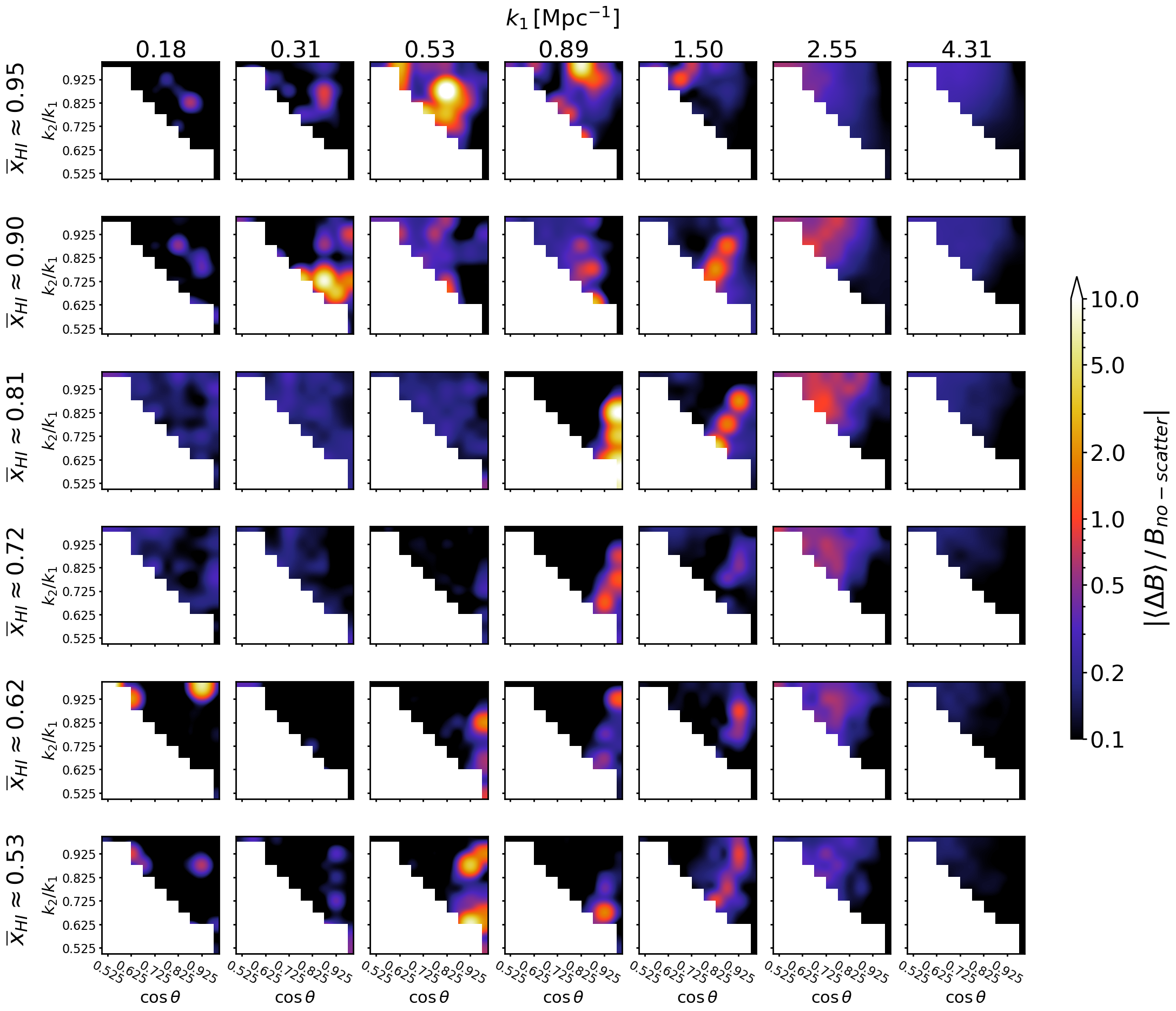}
    \caption{Impact on bispectrum due to the astrophysical scatter is shown here at $z=7.4$. The rows represent the average neutral fractions for a given averaged realization. The columns represent $k_1$ modes for the bispectrum. The color bar scale represent the change ($\langle B \rangle$) compared to the bispectrum for the no-scatter case ($B_{\text{no-scatter}}$).}
    \label{fig:bispec_impact}
\end{figure*}

We present our results on the impact of scatter on the \hi bispectrum in Figure~\ref{fig:bispec_impact}. The labels in Figure~\ref{fig:bispec_impact} are identical to those of Figure~\ref{fig:bispec_scatter}. The quantity plotted in Figure~\ref{fig:bispec_impact} is the ratio, $|\langle \Delta B \rangle/B_{\text{no-scatter}}|$, which quantifies the deviation in bispectrum arising from the astrophysical scatter being taken into account in our reionization model. We see that across the extensive range of length scales investigated ($k_1$ values), the impact due to the scatter varies within a large dynamic range of magnitude. The regions of the bispectrum configuration space where the impact is seen to be high in magnitude are randomly spread for $k_1 \lesssim 1.5$ Mpc$^{-1}$, without any consistent pattern. Here, one can suspect that these random patterns arise mostly due to the statistical variance of the astrophysical scatter in the bispectrum. The magnitude of the impact can be more than a factor of $10$ as well for some of the bispectrum triangle configurations. However, as we quantify in the following subsection, most of the changes are not statistically significant for $k_1$ modes up to 1.5 Mpc$^{-1}$ and can vary substantially from realizations to realizations.\par
At scales $k_1 \sim$ $2.55$ Mpc$^{-1}$, we see that most of the $\cos\theta$ - $k_2/k_1$ bispectrum configuration space is sensitive to the astrophysical scatter, and these additional changes follow a consistent pattern across all the neutral fractions, unlike the intermediate and large scales. The ratio $|\langle \Delta B \rangle/B_{\text{no-scatter}}|$ is $\sim 0.2$ at $\overline{x}_{\rm HI}\approx0.53$, for a significant region of the bispectrum configuration space. At $\overline{x}_{\rm HI}\approx0.81$, the ratio goes up to $\sim 1$, which then declines to $ 0.2 \lesssim |\langle \Delta B \rangle/B_{\text{no-scatter}}| \lesssim 1.0$  again at $\overline{x}_{\rm HI}\approx0.95$. The region of the $\cos\theta$ - $k_2/k_1$ configuration space sensitive to the signatures of astrophysical scatter gradually grows, as we go to higher neutral fractions, and saturation appears at the highest neutral fraction, where the magnitude of the impact is more uniform. In this case, at $\overline{x}_{\rm HI} > 0.8$, enough number of ionized bubbles might not have been formed in the \hi maps that would reflect the signatures of astrophysical scatter. We argue in the next subsection (subsection~\ref{stat_sig}) and Figure~\ref{fig:scatter_ionized_bubbles}, that the signatures of astrophysical scatter stem from the variations in sizes of small ionized bubbles across the multiple realizations. At higher neutral fractions, a lack of sufficient ionized bubbles will reduce the signatures of astrophysical scatter as compared to the lower neutral fractions. Therefore, the impact is seen to peak somewhere at an intermediate neutral fraction.\par
\begin{figure*}
    \centering
    \includegraphics[width=\textwidth]{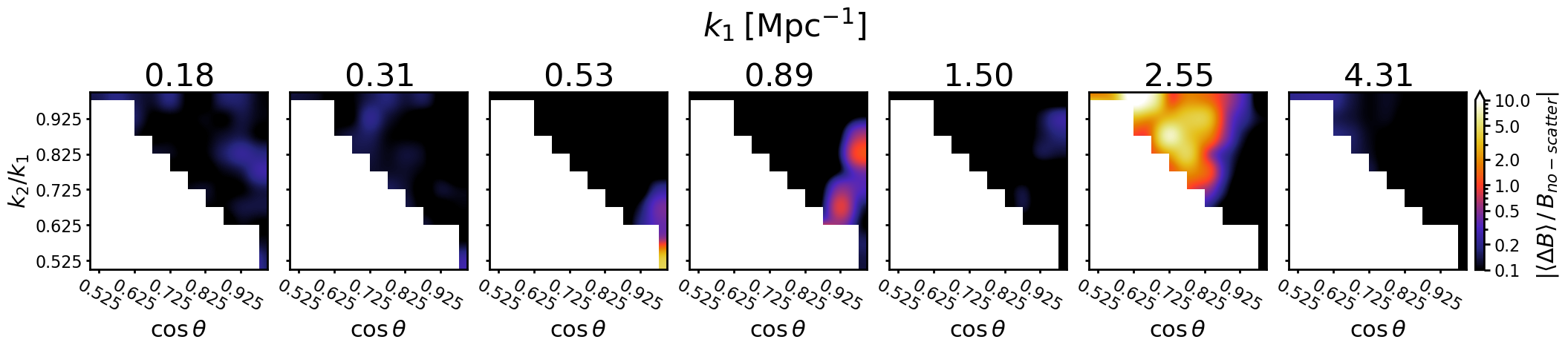}
    \caption{This figure shows the impact of astrophysical scatter for the case at $z=10$ and $\overline{x}_{\rm HI} \approx 0.8$.}
    \label{fig:impact_of_scatter_z10}
\end{figure*}

\label{impact_on_unique_tri}
We also estimate the impact of the scatter at $z=10$ assuming $\overline{x}_{\rm HI} \approx 0.8$, which is shown in Figure~\ref{fig:impact_of_scatter_z10}. Similar to the case at $z=7.4$, we find that the impact of scatter is significantly high at $k_1 \sim 2.55$ Mpc$^{-1}$. We also note that at  $z=10$, the magnitude of the impact is much higher compared to that at $z=7.4$ for the same neutral fraction. At higher redshifts, we expect the number of ionizing sources to be lesser than that at lower redshifts. This means that the number of ionized bubbles will also be less at higher redshifts, and therefore slightly larger than at lower redshifts when compared at the same neutral fraction. This causes the impact of scatter to be more prominent at higher redshifts.

\subsection{Statistical significance}
\label{stat_sig}
Since the astrophysical scatter is a stochastic phenomenon, any statistic used to quantify the impact due to scatter will have an associated variance arising from the stochasticity. It is thus important to investigate the statistical significance of the impact of the astrophysical scatter discussed in the previous subsection~\ref{change_scatter} and Figure~\ref{fig:bispec_impact}. The realization-to-realization variance arising from the astrophysical scatter is estimated for the bispectrum as $\sigma^2_{\Delta B} = \sum_{\text{R}}(\Delta B_{\text{R}} - \langle\Delta B\rangle)^2/N$, with $N=50$ being the number of independent realizations of the scatter. $\Delta B_{\rm R} = B_{\rm R} - B_{\rm no-scatter}$ represents the deviation in the bispectrum for a single realization, arising from the astrophysical scatter compared to the bispectrum without the impact of scatter,  at a fixed neutral fraction. We use the quantity $|\langle\Delta B\rangle/\sigma_{\Delta B}|$ to estimate the statistical significance of the impact of scatter.\par
\begin{figure*}
    \centering
    \includegraphics[width=\textwidth]{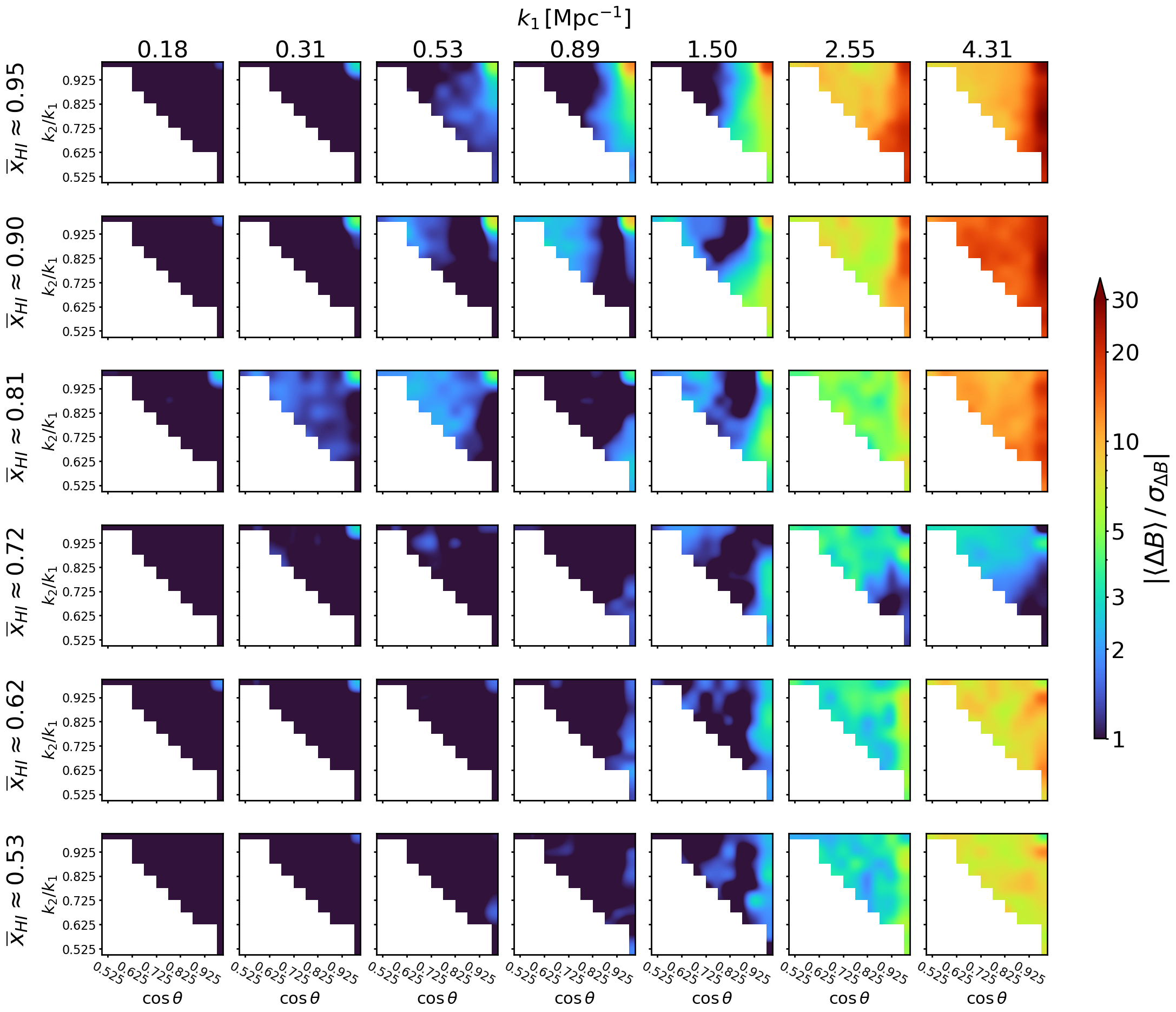}
    \caption{Statistical significance of the impact of scatter on the bispectrum is shown here at $z=7.4$, with rows representing average neutral fraction $\overline{x}_{\rm HI}$ and columns representing $k_1$ mode of the bispectrum.}
    \label{fig:bispec_stat_sig}
\end{figure*}

In Figure~\ref{fig:bispec_stat_sig}, we see that at  large and intermediate scales up to $k_1 \lesssim 1 $ Mpc$^{-1}$ , $|\langle\Delta B\rangle/\sigma_{\Delta B}|\sim 1$, meaning that the impact is not statistically significant. At these scales, there are occasional occurrences of $2\sigma$ statistical significance at $\overline{x}_{\rm HI} \gtrsim 0.81$. These results suggest that the changes in the bispectrum due to astrophysical scatter are not statistically significant at large and intermediate scales.  However, at $k_1 \gtrsim$ 2.55 Mpc$^{-1}$, we see that the changes arising from the astrophysical scatter are statistically significant with more than $3\sigma$ significance at $\overline{x}_{\rm HI}\approx0.53$, for almost all $\cos\theta$ - $k_2/k_1$ configurations of the bispectrum. The statistical significance becomes $\gtrsim 5\sigma$ for $\overline{x}_{\rm HI}\gtrsim0.81$, and further increases for higher neutral fractions. We notice that at $\overline{x}_{\rm HI} \sim 0.7$, a comparatively smaller area in the bispectrum configuration space is statistically significant; however the exact reason for this behavior is not clearly known. A more detailed investigation is required which is deferred for future work.\par

\begin{figure*}
    \centering
    \includegraphics[width=\textwidth]{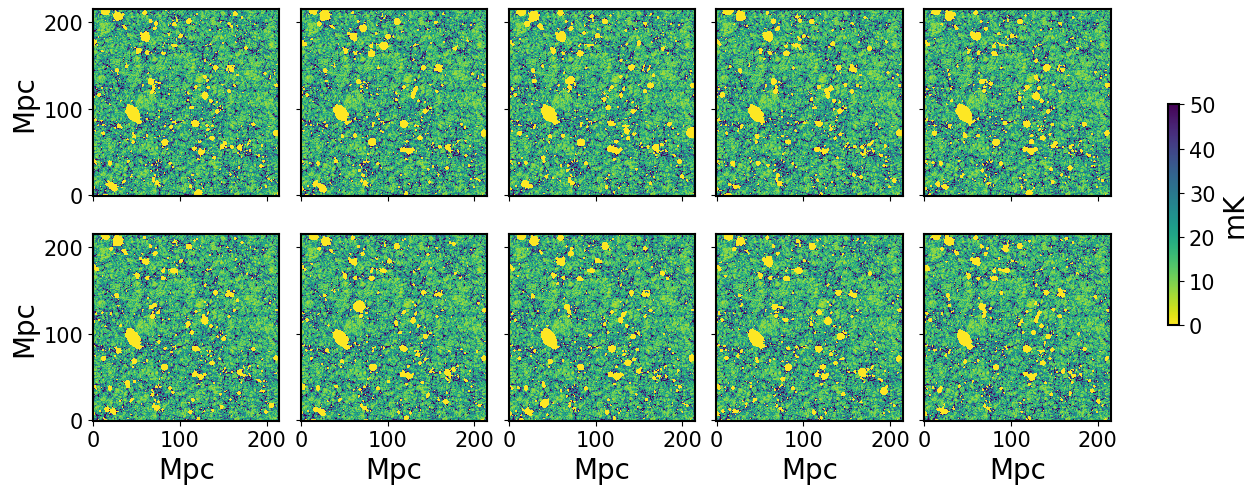}
    \caption{Different  scatter realizations of the full $215\,\text{Mpc}\times215\,\text{Mpc}$ slice is shown for $z=7.4$ at a fixed neutral fraction of $\overline{x}_{\rm HI}\approx 0.81$.}
    \label{fig:scatter_ionized_bubbles_largea_scales}
\end{figure*}

\begin{figure*}
    \centering
    \includegraphics[width=\textwidth]{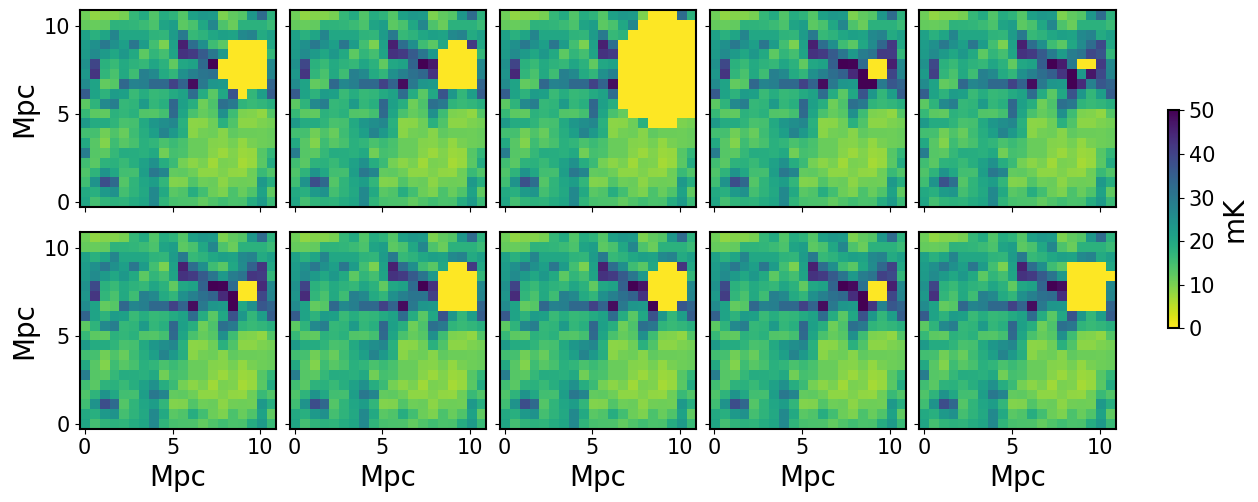}
    \caption{Realization to Realization fluctuation of a particular ionized bubble is shown here on a scale of $\approx 10\,\text{Mpc}\times10\,\text{Mpc}$ for $z=7.4$ at a fixed neutral fraction of $\overline{x}_{\rm HI}\approx 0.81$.}
    \label{fig:scatter_ionized_bubbles}
\end{figure*}

We try to visually understand the impact of the astrophysical scatter being dominant at the small scales compared to the large scales. It is found that large ionized bubbles remain very similar in different realizations of astrophysical scatter. This can be seen in Figure~\ref{fig:scatter_ionized_bubbles_largea_scales} which shows \hi maps for $10$ different realizations of the scatter, whereas the underlying dark-matter field and halo list are the same. We see that the largest ionized bubble in different maps remains largely unaffected. However, the sizes of small ionized bubbles vary considerably in different realizations.  This is more clear in Figure~\ref{fig:scatter_ionized_bubbles} which shows a zoomed-in version of a particular region of Figure~\ref{fig:scatter_ionized_bubbles_largea_scales}, focusing on a small ionized bubble.  Small ionized bubbles encompass a few low-mass dark-matter halos and the number of ionizing photons emitted by them vary in different realizations due to the astrophysical scatter. This results in the size of small ionized bubbles to vary considerably across realizations. This is not the case for large bubbles which encompass many more dark-matter halos and the collective number of ionizing photons contributed by many reionizing sources does not vary much across realizations.  Therefore, the impact of scatter is more prominent in the small ionized bubble size distribution, whereas it is largely negligible and washed away in large ionized bubbles.  In the absence of large ionized bubbles in a highly neutral IGM, at the early stages of reionization, the signatures of astrophysical scatter are retained and captured in the \hi bispectrum at the small scales, with high statistical significance.  In contrast, the signatures of astrophysical scatter are not visible at comparatively larger scales, at the later stages of EoR.

\begin{figure*}
    \centering
    \includegraphics[width=\textwidth]{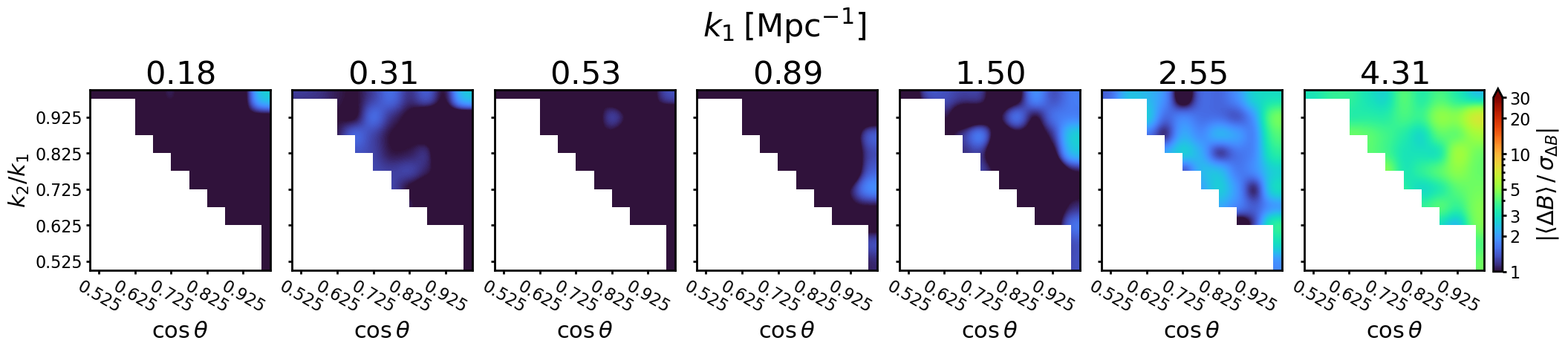}
    \caption{This figure shows the statistical significance of the impact of astrophysical scatter at $z=10$ and $\overline{x}_{\rm HI} \approx 0.8$.}
    \label{fig:stat-sig_scatter_z10}
\end{figure*}

The statistical significance for the impact of astrophysical scatter for $z=10$ and $\overline{x}_{\rm HI} \approx 0.8$ is shown in Figure~\ref{fig:stat-sig_scatter_z10}. At $k_1 \sim 2.55$ Mpc$^{-1}$, where the impact is found to be high, the statistical significance of the impact is around $\sim 3\sigma$. This is lesser than that found at $z=7.4$, where the statistical significance was around $5\sigma$ at the same neutral fraction of $\overline{x}_{\rm HI} \approx 0.8$. We note that since the number of ionized bubbles is lower at higher redshifts, this might somewhat reduce the overall statistical significance of the impact. However, this impact is still statistically significant at a level of $3\sigma$.

\subsection{Sign flip of the [H {\footnotesize I}] 21cm bispectrum}
\begin{figure*}
    \centering
    \includegraphics[width=\textwidth]{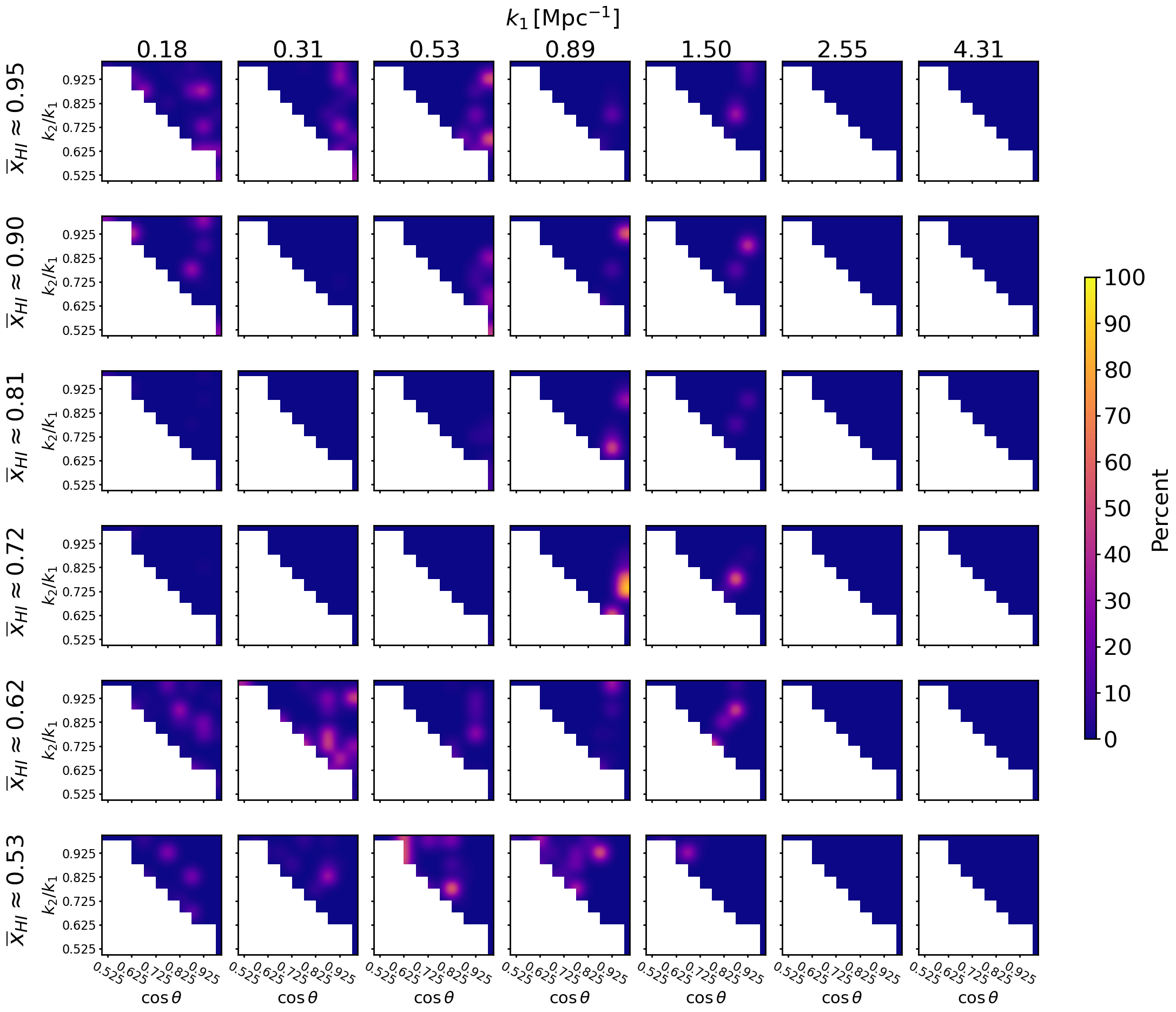}
    \caption{This figure shows the sign flip induced by astrophysical scatter at $z=7.4$ for multiple neutral fractions.}
    \label{fig:bispec_sign-flip_z7.4}
\end{figure*}

\begin{figure*}
    \centering
    \includegraphics[width=\textwidth]{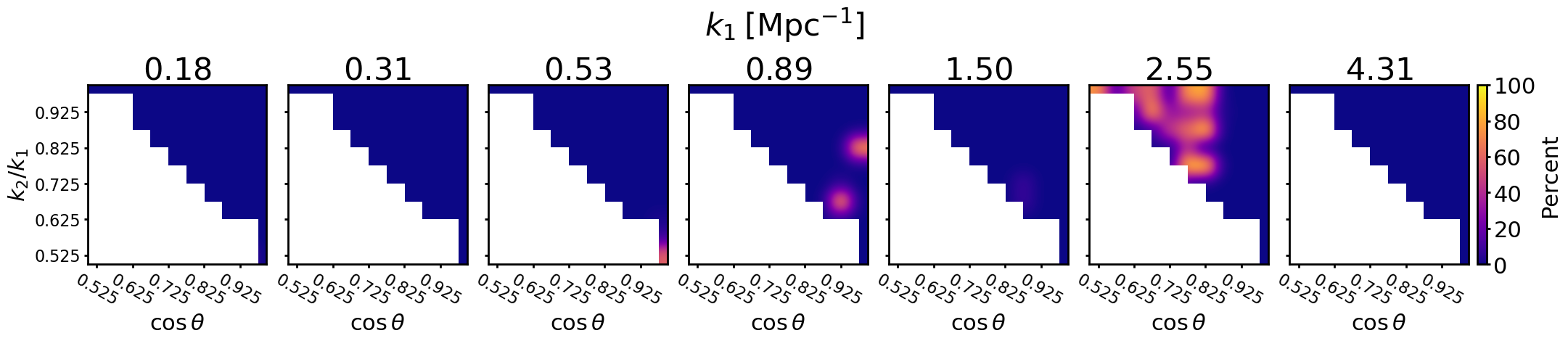}
    \caption{This figure shows the frequency of occurrence of sign flip in the \hi bispectrum for $z=10$ at $\overline{x}_{\rm HI} \approx 0.8$ arising from astrophysical scatter.}
    \label{fig:bispec_sign-flip_z10}
\end{figure*}

We discuss the possibility of a sign flip in the bispectrum induced by the impact of astrophysical scatter. We examine whether the condition $B_{\rm scatter}/B_{\rm no-scatter} < 0$ holds in each cell of the $k_1/k_2-\cos \theta$ space to check for the occurrence of sign flip in the bispectrum. It is repeated for all 50 realizations of the astrophysical scatter for the multiple neutral fractions at $z=7.4$ and for $\overline{x}_{\rm HI} \approx 0.8$ at $z=10$. The occurrence of the sign flip in the \hi bispectrum is counted in all $\sim$50 realizations and the total number of occurrences of this sign flip in each cell of $k_1/k_2-\cos \theta$ space is estimated in percentage. Therefore, a $100$ percent would mean that the sign flips in \hi bispectrum occur in a particular cell in $k_1/k_2-\cos \theta$ space in every realization of the astrophysical scatter. In Figure~\ref{fig:bispec_sign-flip_z7.4}, the sign flip due to scatter for $z=7.4$ is shown for multiple neutral fractions and length scales.
We find that for $z=7.4$ at specific triangle configurations, the frequency of sign flip occurrence is much less than 50 percent in the \hi bispectrum, across multiple neutral fractions and for length scales $k_1 \lesssim 1.5$ Mpc$^{-1}$. It means that for most of the scenarios, there is no sign-flip in the bispectrum, and the fiducial no-scatter model falls in the majority sign. For $z=10$ at $\overline{x}_{\rm HI} \approx 0.8$, a significant area in the bispectrum triangle configuration space at $k_1 \sim 2.55$ Mpc$^{-1}$ shows sign flip occurrence with frequency significantly greater than 50 percent as can be seen in Figure~\ref{fig:bispec_sign-flip_z10}.
Therefore, in this case for most of the scenarios, the \hi bispectrum can flip its sign and the fiducial no-scatter model falls in the minority sign.

\subsection{Comparison with the power spectrum}
\label{pow_spec}
\begin{figure*}
    \centering
    \includegraphics[width=\textwidth]{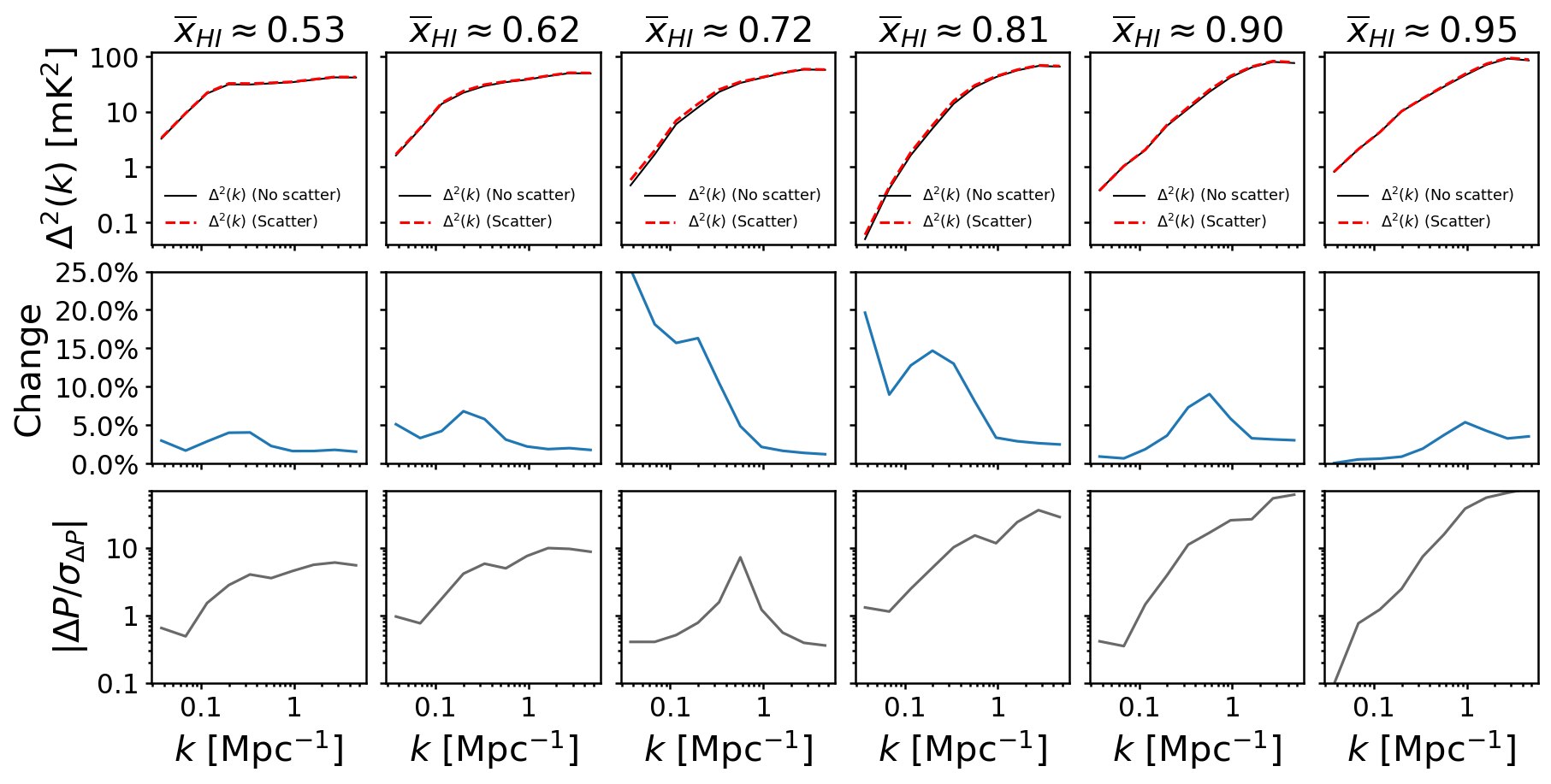}
    \caption{Impact on the power spectrum due to the astrophysical scatter is shown here, along with the statistical significance for multiple neutral fractions at $z=7.4$.}
    \label{fig:powerspectrum_impact}
\end{figure*}

In this section, we do a similar analysis of the impact of astrophysical scatter on the power spectrum, as is done with bispectrum (as described in previous sections), and compare it with the latter to understand how the bispectrum can capture information that the power spectrum might fail to do. We estimate the power spectrum for the no-scatter scenario and all of the $\sim 300$ realizations of astrophysical scatter, with $\sim 50$ 
realizations for each neutral fraction at $z=7.4$. In Figure~\ref{fig:powerspectrum_impact}, top panel, we compare the power spectrum for the scenario with astrophysical scatter, averaged over all realizations (red-dashed line) with that when the scatter is not taken into account (black-solid line). We note that at $\overline{x}_{\rm HI} \sim 0.8$ the large-scale power spectra exhibit a dip compared to the other neutral fractions. Although the density fluctuations are constant (since the redshift is kept fixed), the cross-correlation between the density of the neutral hydrogen and the total overdensity will vary with different neutral fractions. The cross-power spectrum of these two fields is negative when reionization is in its early stage since the overdense regions (preferred location of ionizing sources) get ionized first in our inside-out reionization model. This cross-term contributes to the \hi signal power spectrum and suppresses it at large scales compared to the higher neutral fractions. However, when the ionized bubbles are sufficiently large and numerous at a particular reionization stage, this cross-correlation becomes weaker, and its contribution to the \hi power spectrum goes down. However, the contribution of the neutral hydrogen power spectra to \hi signal power spectra will dominate over this cross-correlation. It will cause the large-scale power spectra to increase. Therefore, at a particular neutral fraction stage, we see the large-scale power spectra have a dip. This phenomenon is consistent with several earlier studies of reionization with radiative transfer and semi-numerical simulations~\cite{Lidz_2007, Mao_2012, Majumdar_2013, Majumdar_2016}. In the middle panel, we present the percentage change in the power spectrum due to scatter and the corresponding statistical significance of the impact in the bottom panel.\par
For the neutral fractions, $\overline{x}_{\rm HI} = 0.53 - 0.81$, the magnitude of the impact ranges from $5 - 15$ per cent, with statistical significance ranging from $4\sigma - 5\sigma$. At $\overline{x}_{\rm HI} = 0.90$ and $\overline{x}_{\rm HI} = 0.95$, the statistical significance is very high ($10\sigma$ and $14\sigma$ respectively), where the magnitude of the impact of astrophysical scatter reaches its peak. However, the magnitude of this impact is less ($\sim 10$ percent and $\sim 5$ percent respectively) compared to the magnitude of the impact at $\overline{x}_{\rm HI} = 0.81$. Unlike the bispectrum, the impact on the power spectrum is at a single-length scale, with the maximum magnitude of the impact at $k \sim 0.2$ Mpc$^{-1}$ being $\lesssim 15$ percent for $\overline{x}_{\rm HI} = 0.81$. The maximum magnitude of the impact declines and shifts to smaller scales for the higher neutral fractions. The maximum magnitude of the impact we find for the neutral fractions considered is broadly consistent with the findings of \cite{Hassan_2022}. However, they present their results based on the ionization power spectrum (which is different from the power spectrum of the brightness temperature fluctuations). They also do not include a corresponding statistical significance of the changes in the power spectrum that they find, arising from the astrophysical scatter.\par
On the other hand, the bispectrum, being a 3-pt Fourier statistic, captures the impact of scatter on the correlations between different length scales. This impact consistently exceeds 20 percent across all neutral fractions ($\overline{x}_{\rm HI}$) $\gtrsim 0.53$ at $k \sim 2.55$ Mpc$^{-1}$, with statistical significance consistently equal to or higher than $3\sigma$, at those scales. At $\overline{x}_{\rm HI} = 0.81$, the impact of astrophysical scatter reaches $\sim 100$ percent for a significant region of the bispectrum triangle configuration space at $5\sigma$ statistical significance. It suggests that the impact of the astrophysical scatter on the \hi signal is captured and characterized in a more detailed manner with the bispectrum as opposed to the power spectrum, which misses out on the signatures of the impact of scatter at the relevant neutral fractions and length-scales, with sufficient magnitude.

\subsection{Detectability of the [H {\footnotesize I}] 21cm auto-bispectrum}
\label{bispec_detect}
\begin{figure*}
    \centering
    \includegraphics[width=\textwidth]{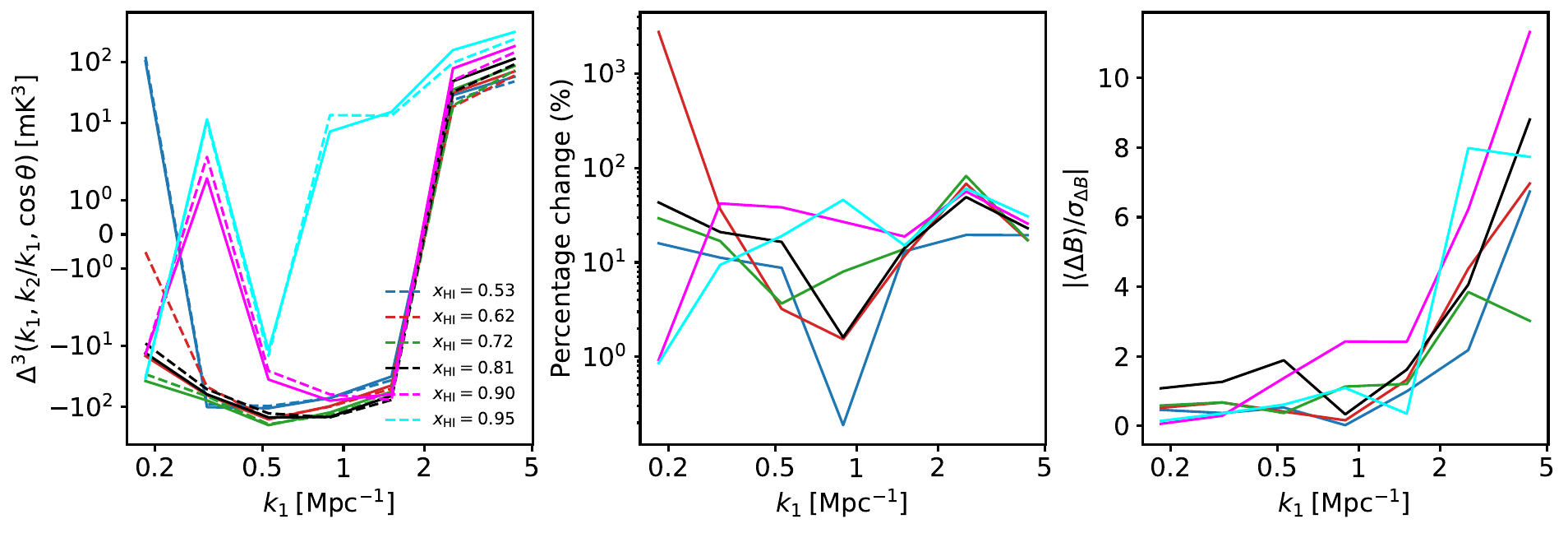}
    \caption{Left panel: The equilateral bispectrum without (dashed lines) and with  (solid lines) the astrophysical scatter is shown here for multiple neutral fractions at $z=7.4$. Middle panel: The impact due to scatter on the equilateral bispectrum is shown here. Right panel: The corresponding statistical significance of the impact of scatter is shown here.}
    \label{fig:bispec_eq}
\end{figure*}

Here, we explore the possibility of detecting the impact of the scatter on the \hi bispectrum considering observations with the planned SKA1-Low. As a test case, we limit our analysis to only equilateral triangle configurations of the bispectrum, which corresponds to the top-left corner of the bispectrum configuration space shown in the previous figures. The equilateral bispectrum is expected to be affected the most by the system noise. Although it is shown in previous studies~\citep{Mondal_2021}, that the squeezed limit bispectrum has the best signal-to-noise ratio for detection, we find that this bispectrum triangle configuration is not likely to be affected by astrophysical scatter. The bispectrum for the equilateral triangle configuration is found to be significantly affected by scatter. Therefore, we focus on the prospects of detectability of the equilateral bispectrum. In Figure~\ref{fig:bispec_eq}, the equilateral bispectrum is shown without (dashed lines) and with (solid lines) the astrophysical scatter, along with the corresponding impact and the statistical significance of the impact, for various neutral fractions at $z=7.4$. It is seen that, except $\overline{x}_{\rm HI} \sim 0.5$, the impact on the equilateral bispectrum is statistically significant with $|\langle\Delta B\rangle/\sigma_{\Delta B}| \gtrsim 3\sigma$ for $k_1 \sim 2.55$ Mpc$^{-1}$.
\par
\begin{figure*}
    \centering
    \includegraphics[width=0.5\textwidth]{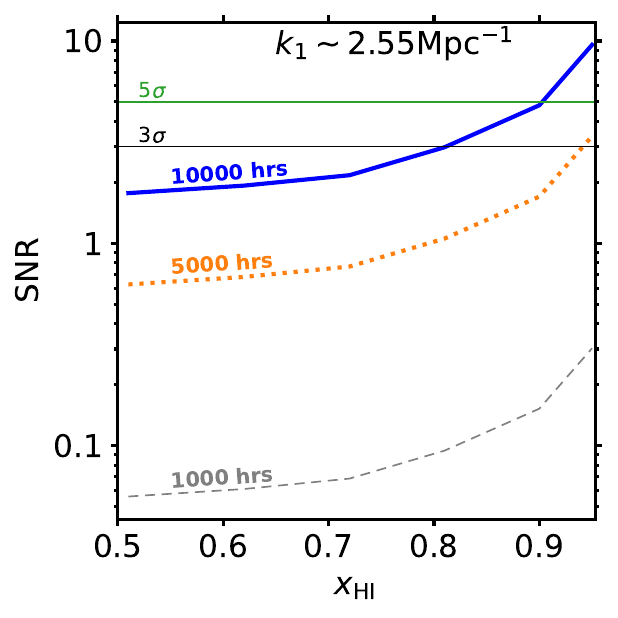}
    \caption{The signal-to-noise ratio of the detection of the equilateral bispectrum at $z=7.4$ is shown, for the scales where the impact of scatter is expected to be significant ($k_1 \sim 2.55$ Mpc$^{-1}$) and considering observations with SKA1-Low for various durations of observational time. The horizontal solid black and solid green line denotes the threshold for $3\sigma$ and $5\sigma$ detection significance respectively.}
    \label{fig:bispec_snr}
\end{figure*}

We compute the variance in the bispectrum ($\sigma^2_{\rm N}(B)$) due to system noise using \citep{Scoccimarro_2004, Liguori_2010}:
\begin{equation}
    \sigma^2_{\text{N}} (B) \approx s_{B}\frac{V_{f}}{V_{B}}P_{\text{N}} (k_1,z)P_{\text{N}} (k_2,z)P_{\text{N}} (k_3,z)
    \label{eq:bispec_noise}
\end{equation}
Here, $V_{f} = (2\pi)^3/V_{\text{s}}$ is the volume of the fundamental cell in the Fourier domain, with $V_{s}$ being the survey volume, and $P_{\rm N}(k, z)$ is the noise power spectrum contributed by the system noise. In equation \ref{eq:bispec_noise}, $s_B$ = 6 for equilateral triangles, and $V_B\approx8\pi^2k_1k_2k_3\Delta k_1\Delta k_2\Delta k_3$. The noise power spectrum due to the system noise in radio interferometric experiment is given by \citep{Bull_2015, Obuljen_2018}:
\begin{equation}
    P_{\text{N}} (k,z) = \frac{T^2_{\text{sys}}(z)\chi^2(z)r_{\nu}(z)\lambda^4(z)}{A^2_{\text{eff}}t_{\text{obs}}n_{\text{pol}}n(\bm{u},z)\nu_{\text{21cm}}}.
\end{equation}
$\chi (z)$ is the comoving distance to redshift $z$, and $r_{\nu}=(c/H(z))(1+z)^2$. $\lambda (z)=21\times(1+z)$~cm and $\nu_{\text{21cm}}=1420\,\text{MHz}$, are the redshifted wavelength and the rest frame frequency of the \hi emission, respectively. We assume the number of polarization ($n_{\text{pol}}$) to be $2$ and $n(\bm{u},z)=N^2_{\text{a}}/(2\pi u^2_{\text{max}})$ is the baseline density which we assume to be constant within the core radius. $N_{\text{a}}$ is the total number of antennae in the experiment and $u_{\text{max}}$ is the maximum baseline in units of $\lambda (z)$. The system temperature $T_\text{sys}$ is modeled as $T_{\text{sys}}(\nu) = 100\, +\, 300(150\,\text{MHz}/\nu)^{2.55}\,\text{K}$ following \cite{Mellema_2013}. $A_{\text{eff}}$ is the effective collecting area of each antenna which is modeled as, $A_{\text{eff}} = A_{\text{eff}}(\nu_{\text{crit}})\times\epsilon(\nu)$~\citep{Bull_2015}, where $\epsilon(\nu)$ is defined as
\begin{equation}
\epsilon (\nu) = \bigg\{
\begin{tabular}{lr}
    $(\nu_{\text{crit}}/\nu)^2$, & $\nu > \nu_{\text{crit}}$ \\
    1, & $\nu \leq \nu_{\text{crit}}$. \\
\end{tabular}
\end{equation}
$A_{\text{eff}}(\nu_{\text{crit}})$ is taken to be 962~m$^2$ at $\nu_{\text{crit}}=110$ MHz~\citep{Giri_2018}. We take $N_{\text{a}}=296$ within a core radius of $R_{\text{max}}=2\,\text{Km}$~\cite{Mazumder_2022}. The relative $k$-bin size is taken as $\Delta k/k\sim1$. We assume different scenarios where the bandwidth is kept fixed at $16$ MHz and vary the observational duration, $t_{\rm obs}$, to estimate the detectability of the equilateral bispectrum for the signal model considered here. We consider three scenarios, with the first being to observe for a total of 1000 hours. In the other two scenarios, we assume that the observation takes 1000 hours per year after SKA1-Low is operational and the observational campaign is carried out for the next couple of years. We assume that this campaign lasts five and ten years, with the total observational time accumulated to 5000 and 10000 hours, respectively. We restrict this exercise to only the scales of our interest, where the impact of scatter is statistically significant and sufficient in magnitude, i.e., $k_1 \sim 2.55$ Mpc$^{-1}$. In Figure~\ref{fig:bispec_snr}, we show the resulting signal-to-noise ratio, considering these three scenarios of observational duration for various neutral fractions that we have explored in this study.\par
We see that in the first scenario, with a total of 1000 hours of observation (grey-dashed line), the equilateral bispectrum is not detectable with a sufficient signal-to-noise ratio. In the case, where the observational campaign lasts for five years with 1000 hours per year (orange-dotted line), the same cross the $1\sigma$ detection limit at $\overline{x}_{\rm HI} \sim 0.8$ and it reaches $3\sigma$ for $\overline{x}_{\rm HI} \sim 0.9$. However, if we adopt a more optimistic case, where the campaign lasts ten years with 1000 hours of observation time per year (blue-solid line), then the equilateral bispectrum is above the $1\sigma$ detection limit for all the neutral fractions at $k_1 \sim 2.55\, \rm Mpc^{-1}$, starting from $\sim 2\sigma$ detection significance at $\overline{x}_{\rm HI} \sim 0.5$. However, at $\overline{x}_{\rm HI} \sim 0.8$ and 0.9, the detection significance is $\sim 3\sigma$ and $\sim 5\sigma$ respectively.\par
The results for the detectability that have been presented here are for a single redshift ($z=7.4$). However, at higher redshifts, one can expect that the signal-to-noise ratio will be further degraded, due to increased system noise temperature $T_{\rm sys}$. We estimated the signal-to-noise ratio of the detectability of the bispectrum for $z=10$ at $\overline{x}_{\rm HI} \approx 0.8$, which was done for the triangle configurations near the vicinity of the equilateral triangles, where statistical significance is high. The signal-to-noise ratio is far below the unity even for $t_{\rm obs}=10000$ hours.\par
We highlight that for higher-order Fourier statistics such as bispectrum, the contributions of cosmic variance to the total bispectrum uncertainty can be significant at large scales. In~\cite{Mondal_2021}, the contribution of the cosmic variance to the total bispectrum uncertainty budget has been studied thoroughly for a range of length scales. The uncertainty contributed by cosmic variance decreases with decreasing length-scale and this is true for the equilateral bispectrum as well, as found by~\cite{Mondal_2021}. They studied a case with roughly $\mathbf{\sim}$1000 hours of observation with SKA1-Low. They found that the contribution from the cosmic variance to the total uncertainty budget dominated at $k_1 \lesssim 0.5\, \rm Mpc^{-1}$ in this case. However, one might expect that cosmic variance in the bispectrum can still contribute to the total signal-to-noise ratio budget down to the scales at $k_1 \sim 2.55\, \rm Mpc^{-1}$ to some extent, and affect the signal-to-noise ratio of the detection at those scales. Also, the estimation of the signal-to-noise ratio of the detection of the equilateral bispectrum, presented here, is analytic, and a full numerical approach~\citep{Mondal_2021} is needed for a more thorough investigation. Therefore, one needs to simulate multiple independent realizations of the underlying dark-matter distribution to estimate the uncertainty contribution from cosmic variance in the bispectrum and multiple realizations of \hi noise maps (telescope noise) for a full numerical approach. This thorough approach is a more computationally challenging task and is beyond the scope of the current article. We plan to take up this exercise in future follow-up work.

\section{Summary and Discussion}
\label{sec:summary}
The \hi signal is a promising probe of the Universe during the EoR and can be used to track the evolution of the early IGM and the reionization process. Although the power spectrum can shed light on many important issues, it can not capture the entire information content in the \hi signal as it is highly non-Gaussian. The variation in the ionizing photon emission rates for host halos of a given mass which is referred to as astrophysical scatter, can introduce an additional non-Gaussianity into the \hi signal. In~\cite{Hassan_2022}, the effects of this astrophysical scatter had been studied in the context of cosmic reionization of the IGM, using power spectra of the ionization field. They found that the power spectra are mostly unaffected by the presence of astrophysical scatter. However, the statistical significance of the same has not been thoroughly studied, and the ionization field is not observable, unlike the \hi signal. On the other hand, bispectra can capture some aspects of the non-Gaussian \hi signal. In this work, we study the impact of the astrophysical scatter on the \hi bispectra during the EoR. We simulated \hi maps using a semi-numerical prescription that also incorporates astrophysical scatter and then estimated the fractional change in the bispectra $|\langle\Delta B\rangle/B_{\text{no-scatter}}|$ of the maps due to the scatter. We generated 50 independent realizations of the astrophysical scatter for each of the six neutral fractions that we considered to quantify the statistical significance of the impact of scatter. The statistical significance of the impact is quantified by $|\langle\Delta B\rangle/\sigma_{\Delta B}|$, where $\sigma^2_{\Delta B}$, which is the variance in the $\Delta B$, arising from the independent realizations of astrophysical scatter, for each of the neutral fractions. Here, the analysis is presented for all unique triangle configurations of the bispectrum, for a range of neutral fractions, at a fixed redshift of $z=7.4$ and for $z=10$ at $\overline{x}_{\rm HI} \approx 0.8$. The equilateral bispectrum is one of the triangle configurations, where a significant impact of the scatter is expected. We also explored the prospects for detecting the small-scale ($k_1 \sim 2.55\, \rm Mpc^{-1}$) bispectrum with the planned SKA1-Low. The key findings of this work are:
\begin{itemize}
    \item The large and intermediate scales in the \hi maps ($k_1 \lesssim 1.5$ \rm Mpc$^{-1}$) are largely unaffected due to astrophysical scatter, as captured by the bispectrum. Although, the magnitude of the fractional change in the bispectrum, $|\langle\Delta B\rangle/B_{\text{no-scatter}}|$, is more than a factor of $\sim 10$ in some regions of the $\cos\theta - k_2/k_1$ configuration space of the bispectrum for these length scales, this impact is found to be non-significant, and are a result of statistical noise.
    \item At the small scales ($k_1 \sim 2.55\, \rm Mpc^{-1}$), we find that the impact of astrophysical scatter on the bispectrum is significant. We find that $|\langle\Delta B\rangle/B_{\text{no-scatter}}| \gtrsim 20$ percent at neutral fractions $\overline{x}_{\rm HI} \gtrsim 0.81$ at $z=7.4$ and a significant region of the $\cos\theta - k_2/k_1$ configuration space of the bispectrum, the impact of the astrophysical scatter is maximum, where $|\langle\Delta B\rangle/B_{\text{no-scatter}}| \sim 100$ percent. For $z=10$ at $\overline{x}_{\rm HI} \approx 0.8$, the impact due to scatter is more prominent ($|\langle\Delta B\rangle/B_{\text{no-scatter}}| \gtrsim 10$), although the statistical significance is somewhat less ($\sim 3\sigma$) as compared to the same neutral fraction at $z=7.4$.
    \item The presence of astrophysical scatter primarily affects the small ionized regions of the IGM the most.  On the other hand, large ionized bubbles would be formed due to ionization from the cumulative photon from multiple sources, thus averaging out the signatures of astrophysical scatter, as similarly argued in \cite{Hassan_2022}. Therefore large-length scales would be primarily unaffected by the presence of scatter in the number distribution of the ionizing photons. At higher redshifts, we expect fewer ionizing sources, and hence fewer ionized bubbles, which would be larger compared to the case at lower redshifts at the same neutral fraction. This might result in the impact of scatter being more prominent, however, due to the lower number of ionized bubbles the statistical significance of the impact of scatter can be reduced compared to lower redshifts.
    \item The astrophysical scatter is not found to induce any significant sign flip occurrence in the \hi bispectrum at $z=7.4$. For $z=10$ at $\overline{x}_{\rm HI} \approx 0.8$, the bispectrum is seen to have a significant occurrence of sign flip for a few specific triangle configurations at $k_1 \sim 2.55$ Mpc$^{-1}$. However, most of the triangle configurations do not show a significant occurrence of sign flip.
    \item In the case of the power spectrum, the magnitude of the impact of scatter is not high enough ($\lesssim$ 10 percent) for most of the neutral fractions and length scales. Occasionally, the magnitude of the impact is high ($\gtrsim 15$ percent), however, this is not statistically significant. On the other hand, wherever the impact is statistically significant, the magnitude of the impact is not sufficient enough ($\lesssim 15$ percent). Therefore, the power spectrum does not capture the signatures of astrophysical scatter adequately at the small scales, unlike the bispectrum.
    \item The equilateral bispectra for $z=7.4$ at the small scales could be detected with $\sim 3\sigma$/$5\sigma$ detection significance at $\overline{x}_{\rm HI} \sim 0.8$/$0.9$ if we consider a very optimistic scenario of observing for 1000 hours per year with the SKA1-Low and this observational campaign lasts for ten years. However, at high redshifts, such as $z=10$, the signal-to-noise ratio for the detectability of the \hi bispectrum for triangle configurations in the vicinity of the equilateral triangle is far below unity, at $k_1 \sim 2.55$ Mpc$^{-1}$. The increased system temperature ($T_{\rm sys}$) of the interferometer at higher redshifts can be a contributing factor for the low signal-to-noise ratio. One might note that cosmic variance can significantly contribute to the total uncertainty budget of the bispectrum at large scales, as investigated in~\cite{Mondal_2021}. However, it remains to be seen how much it might contribute to the total uncertainty budget and affect the detection significance at the small scales ($k_1 \sim 2.55\, \rm Mpc^{-1}$), where the impact of astrophysical scatter is found to be significant. We leave this exercise to future work.
\end{itemize}
We want to point out that the entire study was performed considering only one value of the $\sigma$ (defined in equation \ref{eq:scatter}) and a specific model of the astrophysical scatter. Although we use the scatter in the main sequence SFR from~\cite{Speagle_2014}, this is poorly constrained for redshifts in EoR. Recent observations and simulations~\cite{Sun_2023} suggest that during EoR scatter in SFR can also arise from \textit{bursty} SFR, which can result in a different $\sigma$ than what has been used in this work. The impact of the scatter on the IGM \hi signal, as captured by suitable summary statistics (e.g. bispectrum), may vary with different values of the parameter, $\sigma$. To fully understand, the nature of the impact of scatter on the IGM and the cosmic \hi signal, and how it depends on the relevant parameters, one would be required to generate a large number of statistically independent realizations of \hi signal with different values of $\sigma$, and perform a detailed study of the impact. This interesting aspect requires a detailed follow-up study, which we defer for future work.\par
Further, our study has a few more limitations. Throughout our study, we have assumed that the spin temperature ($T_{\text{s}}$) is much higher than the cosmic microwave background radiation temperature ($T_{\gamma}$). This may not be true during the early stages of the EoR.  This assumption will affect the $\delta T_{\text{b}}$ fluctuations since $\delta T_{\text{b}}(\bm{x},z)\propto(1-T_{\gamma}(z)/T_{\text{s}}(\bm{x},z))$ \citep{Bharadwaj_2005}, and Lyman-$\alpha$ coupling and heating of the IGM will play a role in determining these fluctuations. This study assumes that star-forming galaxies drive the entire cosmic reionization process. We have not considered contributions from other sources, such as uniform ionizing background (UIB) originating from active galactic nuclei or X-ray radiation from X-ray binaries or mini-QSOs~\cite{McQuinn_2012, Mesinger_2013, Majumdar_2016}. Here, we also have not modeled the inhomogenous recombination process of the ionized hydrogen, which would affect the ionization morphology of the Universe. It might affect how the astrophysical scatter of the star-forming galaxies affects the reionization process under this cumulative scenario of all possible sources of ionizing photons, contributing to the reionization processes. Also, we have not considered line-of-sight (LoS) anisotropies such as redshift space distortion and light-cone effects in this study, which are inherently present in the observations. Redshift space distortion is expected to distort the ionized bubbles, including the small ionized bubbles, which are significantly affected by the astrophysical scatter. As shown in Figure 2 of~\cite{Mao_2012}, at the high neutral fractions of $\overline{x}_{\rm HI} \gtrsim 0.8$, the redshift space distortion introduces a significant impact ($\gtrsim 40$ percent) on the \hi power spectrum. Similarly, as investigated in~\cite{Majumdar_2020}, the impact of redshift space distortion in the \hi bispectrum can be significant as well ($\sim 200$ percent), at $k_1 \sim 2.37$ Mpc$^{-1}$. Therefore, the presence of redshift space distortion can modulate the signatures of astrophysical scatter, which needs to be investigated further. Similarly, the impact of the light-cone effect on the \hi bispectrum was quantified in \cite{Mondal_2021}, which can also further impact the signatures of astrophysical scatter in the \hi signal. All of these factors when taken into account cumulatively, are interesting avenues for follow-up work, to investigate the more realistic scenario of the impact of astrophysical scatter on the \hi signal, which we plan to take up in the future.\par

\acknowledgments
We thank the reviewer for providing useful feedback which helped to improve this paper. CSM acknowledges funding from the Council of Scientific and Industrial Research (CSIR) via a CSIR-SFR fellowship, under the grant 09/1022(0080)/2019-EMR-I. SM acknowledges financial support
through the project titled “Observing the Cosmic Dawn in Multicolour using Next Generation Telescopes” funded by the Science
and Engineering Research Board (SERB), Department of Science
and Technology, Government of India through the Core Research
Grant No. CRG/2021/004025. KKD also acknowledges financial support from SERB-DST (Govt.
of India) through a project under MATRICS scheme
(MTR/2021/000384). The simulations and numerical analysis presented here have used the computing resources available to the Cosmology with Statistical Inference (CSI) research group at the Indian Institute of Technology Indore (IIT Indore). CSM would also like to thank Samit Pal and Leon Noble for their helpful discussions.\par
This research made use of arXiv\footnote{\href{https://arxiv.org}{https://arxiv.org}} research sharing platform and NASA Astrophysics Data System Bibliographic
Services\footnote{\href{https://ui.adsabs.harvard.edu/}{https://ui.adsabs.harvard.edu/}}. The following softwares have been used: \texttt{NumPy} \citep{Harris_2020}, \texttt{Astropy}\footnote{\href{https://www.astropy.org}{https://www.astropy.org}} \citep{AstropyCollaboration_2022}, \texttt{N-body}\footnote{\href{https://github.com/rajeshmondal18/N-body}{https://github.com/rajeshmondal18/N-body}} \citep{Bharadwaj_2004}, \texttt{FoF-Halo-Finder}\footnote{\href{https://github.com/rajeshmondal18/FoF-Halo-finder}{https://github.com/rajeshmondal18/FoF-Halo-finder}} \citep{Mondal_2015}, \texttt{ReionYuga}\footnote{\href{https://github.com/rajeshmondal18/ReionYuga}{https://github.com/rajeshmondal18/ReionYuga}} \citep{Choudhury_2009, Majumdar_2014, Mondal_2017} and \texttt{DviSukta}\footnote{\href{https://github.com/rajeshmondal18/DviSukta}{https://github.com/rajeshmondal18/DviSukta}} \citep{Mondal_2021}.


\bibliographystyle{JHEP}
\bibliography{ref}
\end{document}